\documentclass{pasj00}

\begin{document}
\SetRunningHead{S. Hasegawa et al.}{Lightcurve Survey of V-type Asteroids}

\title{Lightcurve Survey of V-type Asteroids in the Inner Asteroid Belt}

 \author{%
   Sunao \textsc{Hasegawa},\altaffilmark{1}
   Seidai \textsc{Miyasaka},\altaffilmark{2}
   Hiroyuki \textsc{Mito},\altaffilmark{3}
   Yuki \textsc{Sarugaku},\altaffilmark{1}
   Tomohiko \textsc{Ozawa},\altaffilmark{4}
   Daisuke \textsc{Kuroda},\altaffilmark{5}
   Setsuko \textsc{Nishihara},\altaffilmark{1,6}
   Akari \textsc{Harada},\altaffilmark{6}
   Michitoshi \textsc{Yoshida},\altaffilmark{7}
   Kenshi \textsc{Yanagisawa},\altaffilmark{5}
   Yasuhiro \textsc{Shimizu},\altaffilmark{5}
   Shogo \textsc{Nagayama},\altaffilmark{8}
   Hiroyuki \textsc{Toda},\altaffilmark{5}
   Kouji \textsc{Okita},\altaffilmark{5}
   Nobuyuki \textsc{Kawai},\altaffilmark{9}
   Machiko \textsc{Mori},\altaffilmark{10}
   Tomohiko \textsc{Sekiguchi},\altaffilmark{11}
   Masateru \textsc{Ishiguro},\altaffilmark{12}
   Takumi \textsc{Abe},\altaffilmark{1}
   and
   Masanao \textsc{Abe}\altaffilmark{1}}
 \altaffiltext{1}{Institute of Space and Astronautical Science, Japan Aerospace Exploration Agency, 3-1-1 Yoshinodai, Sagamihara, Kanagawa 252-5210, Japan}
 \email{hasehase@isas.jaxa.jp}
 \altaffiltext{2}{Tokyo Metropolitan Government, 2-8-1 Nishishinjyuku, Shinjyuku, Tokyo 163-8001, Japan}
 \altaffiltext{3}{Kiso Observatory, Institute of Astronomy, The University of Tokyo, 10762-30 Mitake, Kiso, Nagano 397-0101, Japan}
 \altaffiltext{4}{Misato Observatory, 180 Matsugamine, Misato, Wakayama 640-1366, Japan}
 \altaffiltext{5}{Okayama Astrophysical Observatory, National Astronomical Observatory, 3037-5 Honjo, Kamogata-cho, Asakuchi, \\ Okayama 719-0232, Japan}
 \altaffiltext{6}{Department of Earth and Planetary Science, The University of Tokyo, 7-3-1 Hongo, Bunkyo, Tokyo 113-0033, Japan}
 \altaffiltext{7}{Hiroshima Astrophysical Science Center, Hiroshima University, Higashi-Hiroshima, Hiroshima 739-8526, Japan}
 \altaffiltext{8}{National Astronomical Observatory of Japan, 2-21-1 Osawa, Mitaka, Tokyo 181-8588, Japan}
 \altaffiltext{9}{Department of Physics, Tokyo Institute of Technology, 2-12-1 Ookayama, Meguro, Tokyo 152-8551, Japan}
 \altaffiltext{10}{Graduate School of Science, Japan Women's University, 2-8-1 Mejirodai, Bunkyo, Tokyo 112-8681, Japan}
 \altaffiltext{11}{Asahikawa Campus, Hokkaido University of Education, 9 Hokumon, Asahikawa, Hokkaido 070-8621, Japan}
 \altaffiltext{12}{Astronomy Program, Department of Physical and Astronomy, Seoul National University, San 56-1, Sillim-dong, Gwanak-gu, Seoul 151-742, Korea}


%


\maketitle

\begin{abstract}
We have observed the lightcurves of 13 V-type asteroids ((1933) Tinchen, (2011) Veteraniya, (2508) Alupka, (3657) Ermolova, (3900) Knezevic, (4005) Dyagilev, (4383) Suruga, (4434) Nikulin, (4796) Lewis, (6331) 1992 $\mathrm{FZ_{1}}$, (8645) 1998 TN, (10285) Renemichelsen, and (10320) Reiland). 
Using these observations we determined the rotational rates of the asteroids, with the exception of  Nikulin and Renemichelsen. 
The distribution of rotational rates of 59 V-type asteroids in the inner main belt, including 29 members of the Vesta family that are regarded as ejecta from the asteroid (4) Vesta, is inconsistent with the best-fit Maxwellian distribution.
This inconsistency may be due to the effect of thermal radiation Yarkovsky--O'Keefe--Radzievskii--Paddack (YORP) torques, and implies that the collision event that formed V-type asteroids is sub-billion to several billion years in age.
\end{abstract}

\section{Introduction}
Rotational rate is one of the basic physical parameters of asteroids, along with albedo, size, shape, and spin vector.
It is possible to determine the rotational rate of an asteroid from its lightcurve.
The lightcurve of an asteroid can readily be obtained by relative photometric observations with telescopes of small diameter, although this takes considerable time.
Multi-epoch lightcurve observations can be used to derive the shapes and spin vectors of asteroids (\cite{Magnusson1989}; \cite{Kaasalainen2001}).
Combining multi-epoch lightcurve observations with mid-infrared or occultation data enables the albedo and size of asteroids to be determined (\cite{Mueller2007}; \cite{Durech2011}).

The rotation rate of asteroids is a useful means to constrain asteroidal evolution. 
It is known that the rotation rate distribution of asteroids whose diameters are larger than $\sim$40 km is close to a Maxwellian distribution \citep{Pravec2002}.
Numerical simulations have shown that spin rates generated by the collisional history of asteroids generates a Maxwellian distribution \citep{Salo1987}. 
This suggests that such asteroids are planetesimals of the main belt region and/or their remnants produced by collisional disruption.
In contrast, the rotation rate distribution of asteroids smaller than $\sim$40 km in diameter does not follow a Maxwellian distribution.
This suggests that the driving forces for the change in rotation rates results from catastrophic impact, as well as secondary collisions, re-accretion, and Yarkovsky--O'Keefe--Radzievskii--Paddack (YORP) effects.

The asteroid (4) Vesta is considered to be the smallest terrestrial planet as it is the only differentiated asteroid with an intact internal structure with a basaltic surface, an ultramafic mantle, and a metal core. 
The surface spectrum of Vesta has absorption bands in the visible and near-infrared wavelength regions that are similar in spectral shape, wavelength, and depth to the bands found in spectra of howardite, eucrite, and diogenite (HED) meteorites (\cite{McCord1970}; \cite{Larson1975}). 
A vast crater 460 km in diameter and 13 km deep with a central peak and raised rim is present at Vesta's south pole, which is a triaxial ellipsoid with dimensions of 578, 580, and 458 km \citep{Thomas1997}. 
\citet{Zappala1990b} inferred that Vesta has an asteroid family and \citet{Mothe-Diniz2005} reported that the Vesta family contains ca. 4500 members. 
Numerous asteroids with Vesta-like visible spectra are typically referred to as `V-type asteroids'. These asteroids have a near-Earth orbit \citep{Cruikshank1991} in the inner main belt, which is influenced by a 3:1 mean motion resonance and the $\nu$6 secure resonance of Jupiter \citep{Binzel1993}. 
Approximately 3000 V-type asteroids have been identified by spectroscopic and spectrophotometric methods (\cite{Cruikshank1991}; \cite{Xu1995}; \cite{Spahr1997}; \cite{Binzel2001}; \cite{Bus2002}; \cite{Binzel2004a}; \cite{Binzel2004b}; \cite{Duffard2004}; \cite{Lazzaro2004}; \cite{Marchi2005}; \cite{Alvarez-Candal2006}; \cite{deLeon2006}; \cite{Duffard2006}; \cite{Roig2006}; \cite{Davies2007}; \cite{Carvano2010}; \cite{Solontoi2012}; \cite{Sanchez2013}). 
\citet{Gladman1997} used numerical simulations to propose that most impact objects to Earth belong to the Vesta family and \citet{Binzel2004b} noted that the most likely source of V-type near-Earth asteroids is the Vesta family, given the lack of V-type Mars crossers. 
\citet{Carruba2007} showed that V-type asteroids outside the Vesta family were scattered by close encounters with Vesta, and that they formerly belonged to the Vesta family. 
\citet{Roig2012} constrained the mechanisms of a large fraction of V-type asteroids that are observed to be outside the Vesta family.
These studies all suggest that most V-type asteroids in the inner main belt are ejecta from Vesta, and that HED meteorites are sourced from Vesta and V-type asteroids through the 3:1 and $\nu$6 resonances.

The rotational rates of V-type asteroids are poorly constrained, due to the limited number of such asteroids that have currently been studied. As such, in order to better constrain the rotational rate distribution of asteroids sourced from Vesta, we have begun (since Fall 2003) lightcurve observations of V-type asteroids in the inner main belt (semi major axis 2.1 $\lesssim$ a $\lesssim$ 2.5) and determined rotational rates, along with amplitude variations (\cite{Hasegawa2004}; \cite{Hasegawa2008}). 
Our study contributes to determination of an increased number of reliable rotation rates that allows statistically meaningful interpretations of this data for V-type asteroids to be carried out. 

Herein, we present lightcurve observations for a number of V-type asteroids. Subsequent sections of this paper describe the observations and data reduction procedures (Section 2), results of the lightcurve analyses (section 3), and implications of our results for the origin(s) of V-type asteroids (Section 4).

\section{Observations and data reduction procedures}
Our observational survey of lightcurves for V-type asteroids was carried out using six different telescopes between Fall 2004 and Spring 2005. 
Fifty-nine lightcurves were recorded at the Kiso Observatory using two different telescopes in Nagano, Japan (MPC code 381), the Miyasaka Observatory in Yamanashi, Japan (MPC code 366), the Okayama Astrophysical Observatory in Okayama, Japan (MPC code 371), the Misato Observatory in Wakayama, Japan (no MPC code), and the UH88 telescope in Hawaii, USA (MPC code 568). 
Details of the nightly observational conditions are listed in Table \ref{tab:Observational circumstances}.

\begin{longtable}{rlclcccrcc}
  \caption{Observational conditions.\footnotemark[$*$]}\label{tab:Observational circumstances}
  \hline
  NUM & NAME & Type & Date & Duration & $R_{\rm h}$  & $\Delta$  & $\alpha$ & Telescope & Band
\\ 
    &   &   &    &   [hr] &  [AU] &  [AU] &  [$\timeform{D}$] &   & 
\\
\endfirsthead
  \hline
  NUM & NAME & Type & Date & Duration & $R_{\rm h}$ & $\Delta$ & $\alpha$ & Telescope & Band
\\
  \hline
\endhead
  \hline
\endfoot
  \hline
\multicolumn{1}{@{}l}{\rlap{\parbox[t]{1.0\textwidth}{\small
\footnotemark[$*$]The heliocentric distance ($R_{\rm h}$), geocentric distance ($\Delta$), and phase angle ($\alpha$) for observing asteroids were obtained through the JPL HORIZON ephemeris generator system of NASA\footnotemark[1].
}}}
\endlastfoot
  \hline
1933&Tinchen&V&2005.3.7&6.3&2.611&1.632&4.7&1.05 m - Kiso&R\\
1933&Tinchen&V&2005.3.8&4.9&2.611&1.63&4.2&1.05 m - Kiso&R\\
1933&Tinchen&V&2005.3.12&4.2&2.613&1.623&2.4&0.36 m - Miyasaka&R\\
1933&Tinchen&V&2005.3.13&6.4&2.614&1.622&2.0&0.36 m - Miyasaka&R\\
477&Italia&S&2005.3.7&6.3&2.824&1.846&4.3&1.05 m - Kiso&R\\
477&Italia&S&2005.3.8&4.9&2.823&1.842&4.0&1.05 m - Kiso&R\\
477&Italia&S&2005.3.12&4.2&2.820&1.830&2.2&0.36 m - Miyasaka&R\\
477&Italia&S&2005.3.13&6.4&2.819&1.827&1.8&0.36 m - Miyasaka&R\\
2011&Veteraniya&V&2004.11.12&3.6&2.563&2.213&22.4&1.05 m - Kiso&R\\
2011&Veteraniya&V&2005.1.8&8.6&2.633&1.694&7.9&0.36 m - Miyasaka&R\\
2011&Veteraniya&V&2005.1.9&9.1&2.634&1.691&7.5&0.36 m - Miyasaka&R\\
18590&1997 $\mathrm{YO_{10}}$&S&2004.11.12&3.6&2.566&2.213&22.4&1.05 m - Kiso&R\\
22034&1999 $\mathrm{XL_{168}}$&&2004.11.12&3.6&2.589&2.237&22.2&1.05 m - Kiso&R\\
2508&Alupka&V&2004.11.12&8.7&2.461&1.531&10.1&1.05 m - Kiso&R\\
2508&Alupka&V&2004.11.13&4.1&2.462&1.528&9.7&1.05 m - Kiso&R\\
2508&Alupka&V&2004.11.15&8.2&2.464&1.521&8.8&1.05 m - Kiso&R\\
3657&Ermolova&V&2005.5.5&0.5&2.538&1.716&16.1&1.05 m - Kiso&R\\
3657&Ermolova&V&2005.5.9&1.6&2.534&1.749&17.3&1.05 m - Kiso&R\\
3657&Ermolova&V&2005.5.14&2.0&2.529&1.794&18.7&1.05 m - Kiso&R\\
3900&Knezevic&V&2005.1.16&2.1&2.342&1.386&7.3&1.05 m - Kiso&R\\
3900&Knezevic&V&2005.1.17&2.5&2.344&1.384&6.8&1.05 m - Kiso&R\\
3900&Knezevic&V&2005.1.18&1.1&2.346&1.382&6.3&1.05 m - Kiso&R\\
3900&Knezevic&V&2005.2.4&7.5&2.372&1.393&3.8&0.36 m - Miyasaka&R\\
29976&1999 $\mathrm{NE_{6}}$&D&2005.1.16&2.1&5.114&4.169&3.4&1.05 m - Kiso&R\\
29976&1999 $\mathrm{NE_{6}}$&D&2005.1.17&2.5&5.114&4.164&3.2&1.05 m - Kiso&R\\
46121&2001 $\mathrm{FB_{36}}$&C&2005.1.16&2.1&2.891&1.940&6.0&1.05 m - Kiso&R\\
46121&2001 $\mathrm{FB_{36}}$&C&2005.1.17&2.5&2.891&1.935&5.6&1.05 m - Kiso&R\\
4005&Dyagilev&V&2005.3.7&6.7&2.801&1.832&5.5&1.05 m - Kiso&R\\
4005&Dyagilev&V&2005.3.8&3.3&2.800&1.829&5.2&1.05 m - Kiso&R\\
4005&Dyagilev&V&2005.3.13&3.3&2.798&1.817&4.0&0.36 m - Miyasaka&R\\
4383&Suruga&V&2004.12.17&3.2&2.281&1.382&13&1.05 m - Misato&R\\
4383&Suruga&V&2004.12.20&2.1&2.282&1.401&14.1&2.24 m - UH88&R\\
4434&Nikulin&V&2004.10.15&6.2&2.119&1.127&4.0&0.36 m - Miyasaka&R\\
4434&Nikulin&V&2004.10.16&7.7&2.119&1.127&3.6&0.36 m - Miyasaka&R\\
4434&Nikulin&V&2004.10.17&7.7&2.119&1.127&3.4&0.36 m - Miyasaka&R\\
89481&2001 $\mathrm{XH_{27}}$&X&2004.10.16&7.7&1.826&0.834&4.4&0.36 m - Miyasaka&R\\
89481&2001 $\mathrm{XH_{27}}$&X&2004.10.17&7.7&1.826&0.833&4.0&0.36 m - Miyasaka&R\\
4796&Lewis&V&2004.9.15&1.9&1.983&1.033&13.3&1.05 m - Misato&R\\
4796&Lewis&V&2004.10.1&2.6&2.003&1.008&4.4&0.36 m - Miyasaka&R\\
4796&Lewis&V&2004.10.14&3.9&2.021&1.030&4.5&0.30 m - Kiso&R\\
4796&Lewis&V&2004.10.15&1.4&2.022&1.033&5.0&0.30 m - Kiso&R\\
6331&1992 $\mathrm{FZ_{1}}$&V&2004.10.15&3.5&2.171&1.191&6.6&0.30 m - Kiso&R\\
6331&1992 $\mathrm{FZ_{1}}$&V&2004.10.16&1.4&2.172&1.194&6.8&0.50 m - Okayama&R\\
6331&1992 $\mathrm{FZ_{1}}$&V&2004.10.17&5.7&2.173&1.197&7.1&0.50 m - Okayama&R\\
6331&1992 $\mathrm{FZ_{1}}$&V&2004.11.6&1.7&2.200&1.309&14.7&0.36 m - Miyasaka&R\\
8645&1988 TN&V&2005.2.4&3.4&2.471&1.491&3.5&1.05 m - Kiso&R\\
8645&1988 TN&V&2005.2.5&7.4&2.471&1.491&3.2&0.36 m - Miyasaka&R\\
8645&1988 TN&V&2005.2.11&7.0&2.475&1.493&3.1&0.36 m - Miyasaka&R\\
11321&Tosimatsumoto&XL&2005.2.4&2.0&3.255&2.277&2.6&1.05 m - Kiso&R\\
11321&Tosimatsumoto&XL&2005.2.5&7.4&3.255&2.276&2.4&0.36 m - Miyasaka&R\\
10285&Renemichelsen&V&2004.11.12&2.8&2.374&1.449&10.9&1.05 m - Kiso&R\\
10285&Renemichelsen&V&2004.11.13&5.9&2.376&1.455&11.2&1.05 m - Kiso&R\\
10285&Renemichelsen&V&2004.11.15&8.1&2.378&1.470&12.0&1.05 m - Kiso&R\\
10320&Reiland&V&2004.11.12&8.8&2.310&1.328&3.9&1.05 m - Kiso&R\\
10320&Reiland&V&2004.11.13&5.1&2.312&1.329&3.7&1.05 m - Kiso&R\\
10320&Reiland&V&2004.11.15&8.3&2.315&1.331&3.6&1.05 m - Kiso&R\\
10389&Robmanning&&2004.11.15&8.3&1.950&0.966&4.1&1.05 m - Kiso&R\\
10443&van de Pol&S&2004.11.15&8.3&2.028&1.045&4.2&1.05 m - Kiso&R\\
41051&1999 $\mathrm{VR_{10}}$&&2004.11.15&8.3&2.665&1.684&3.2&1.05 m - Kiso&R\\

\end{longtable}
\footnotetext[1]{$\langle$http://ssd.jpl.nasa.gov/horizons.cgi\#top$\rangle$.}

Thirty-three lightcurve observations from the 1.05 m Schmidt telescope at the Kiso Observatory were made using an SITe TK2048E CCD detector (2048 x 2048 pixels), giving an image scale of 1.5\timeform{"}/pixel located in the Schmidt focus. The field of view of the CCD was \timeform{51'} $\times$ \timeform{51'}. Integration times for the observations were 120 to 300 s. The full-width half maximum of star images at this site was typically 5--8\timeform{"}. 

Three lightcurves were recorded using the 0.30 m Dall--Kirkham telescope at the Kiso Observatory (K.3T). K.3T consisted of a MUTOH CV16II CCD camera with a Kodak KAF-1600 chip format, which captures 1536 $\times$ 1024 pixels with an image scale of 1.35\timeform{"}/pixel under 2 $\times$ 2 binning, giving a 17.3\timeform{'} $\times$ 11.5\timeform{'} sky field. The integration times for these observations were 300 s. 

Eighteen of the lightcurves were obtained using the 0.36 m Ritchey--Chr\'etien telescope at the Miyasaka Observatory. The telescope was equipped with an SBIG STL-1001E CCD camera with Kodak KAF-1001E detector whose format is 1024 $\times$ 1024 pixels. This system produced image dimensions of 1.7\timeform{"}/pixel, yielding a field of view of 29.4\timeform{'} $\times$ 29.4\timeform{'}. The observational integration time was 300 s. The full-width half maximum of stellar images at this site was typically 5--8\timeform{"}.

Two lightcurves were made with the 0.50 m Classical--Cassegrain telescope (MITSuME) at the Okayama Astrophysical Observatory \citep{Yanagisawa2010}. An Apogee U6 camera with a Kodak KAF-1001E detector yielded a format of 1024 $\times$ 1024 pixels with an image scale of 1.5\timeform{"}/pixel and a sky field of 25.6\timeform{'} $\times$ 25.6\timeform{'}. Integration times for these two lightcurves were 120 s. The full-width half maximum of stellar images at this site was typically 3--5\timeform{"}.

Two lightcurve observations were carried out with the 1.05 m Cassegrain telescope at the Misato Observatory. The lightcurves were recorded with a SBIG ST-9E CCD camera and Kodak KAF-0261E detector (512 $\times$ 512 pixels; angular resolution of 1.0\timeform{"}/pixel); sky field of \timeform{9'} $\times$ \timeform{9'}). Integration times for these observations were 30 to 180 s. The full-width half maximum of stellar images at this site was typically 3--5\timeform{"}.

One lightcurve was recorded using the 2.24 m telescope at Mauna Kea (UH88), equipped with a wide-field imager comprising eight 2048 $\times$ 4096 pixel CCDs. The projected area was \timeform{33'} $\times$ \timeform{33'}, which corresponds to an angular resolution of 0.45\timeform{"}/pixel. The integration times were 30 s. The full-width half maximum of stellar images at this site was ca. 1\timeform{"}.

All observations were carried out through an R band filter. 
The dark images for correction were constructed from a median combination of 5--20 dark frames. 
Dome flat fielding images were obtained from the 1.05 m Schmidt, 0.36 m Miyasaka, 1.05 m Misato, and UH88 telescopes. 
Flat fielding images by K.3T and MITSuME were taken during the evening and/or morning twilight. 
Stacked flat fielding images for correction were created by a median combination of 10--20 flat fielding frames. 
Light image reduction involved dark subtraction and flat-field correction, and was performed by Image Reduction and Analysis Facility (IRAF) software. 
The asteroids and stellar stars for comparison were measured through a circular aperture with a diameter three times that of the full-width half maximum size used for the APPOT task of IRAF.

All lightcurves were constructed using relative magnitudes.  
The fluxes of 5--20 stars in the same frame as each asteroid were measured for comparison. 
The flux difference between the asteroid and sum of the comparison stars were then computed. 
Variable stars were not used as comparison stars. 

Light travel time was corrected to obtain an accurate lightcurve epoch. 
A G parameter value for V-type asteroids is necessary for phase function correction, and we used values of 0.32 (G parameter value of Vesta) and 0.15 (G parameter value of typical asteroids). 
\citet{Zappala1990a} and \citet{Ohba2003} reported that the amplitude of asteroid lightcurves is affected by the phase angle. 
As such, a coefficient value (m = 0.02) equal to the median value for surface roughness of asteroids \citep{Ohba2003} was adopted to correct lightcurve amplitude for the phase angle = \timeform{0D}.

Fourier analysis (FFT) and phase dispersion minimization methods (PDM) described by \citet{Stellingwerf1978} were utilized for determination of the asteroidal rotational periods.  Rotational periods were finally determined by considering the best result obtained by the two methods.

\section{Results}
\subsection{V-type asteroids}
\subsubsection{(1933) Tinchen}
\begin{figure}
  \begin{center}
    \FigureFile(80mm,80mm){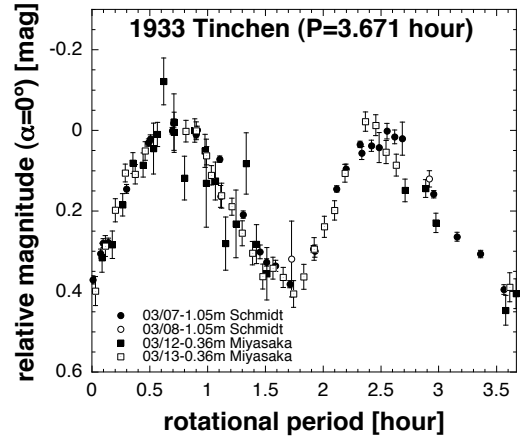}
  \end{center}
  \caption{Composite rotational lightcurve of the asteroid (1933) Tinchen.
}
\label{fig:1933}
\end{figure}

The asteroid (1933) Tinchen is a V-type asteroid \citep{Xu1995} that belongs to the Vesta family \citep{Mothe-Diniz2005}.
\citet{Masiero2011} reported that the diameter of (1933) Tinchen is 6.45 km. 
Tinchen was observed during its apparition in March 2005 (Fig. \ref{fig:1933}), 
and a rotational period of 3.671 $\pm$ 0.002 h was obtained.
On the Observatoire de Geneve website of Behrend (http://obswww.unige.ch/$\sim$behrend/page\_cou.html), the rotational period of (1933) Tinchen was reported to be 3.67 h, which is in excellent agreement with our result.
The quality code (U) scale as used for the Asteroid Lightcurve Database by \citet{Warner2009}. 
The asteroid is valued as U = 3$\mathrm{-}$.

\bigskip
\subsubsection{(2011) Veteraniya}
\begin{figure}
  \begin{center}
    \FigureFile(80mm,80mm){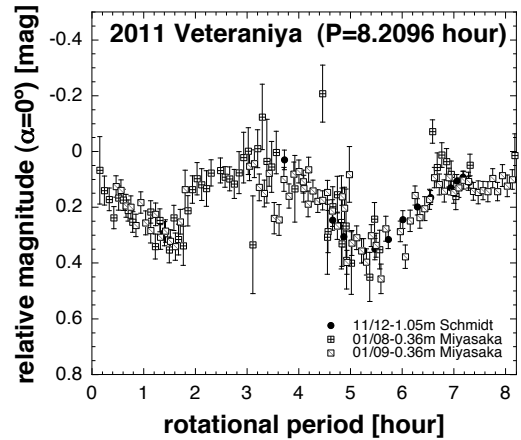}
  \end{center}
  \caption{Composite rotational lightcurve of the asteroid (2011) Veteraniya.
}
\label{fig:2011}
\end{figure}

The asteroid (2011) Veteraniya has a Vesta-like spectrum \citep{Xu1995} and is a member of the Vesta family \citep{Mothe-Diniz2005}. 
The diameter of the asteroid has been estimated to be 5.19 km \citep{Masiero2011}, but the period of this asteroid was not known prior to our study. 
Veteraniya was observed during its apparition in November 2004 and January 2005 (Fig. \ref{fig:2011}), and yielded a rotational period of 8.2096 $\pm$ 0.0003 h.
The asteroid is estimated as U = 3$\mathrm{-}$.

\bigskip
\subsubsection{(2508) Alupka}
\begin{figure}
  \begin{center}
    \FigureFile(80mm,80mm){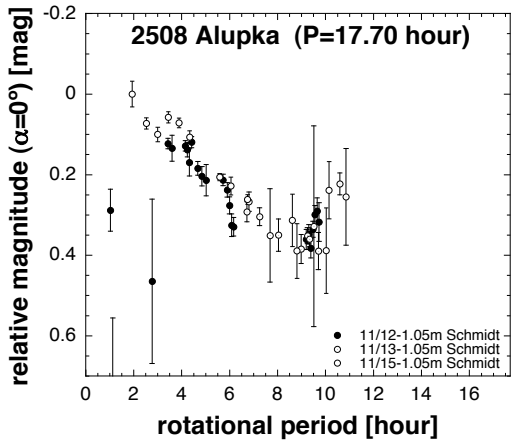}
  \end{center}
  \caption{Composite rotational lightcurve of the asteroid (2508) Alupka.
}
\label{fig:2508}
\end{figure}

The asteroid (2508) Alupka is a Vestoid \citep{Bus2002} and a member of the Vesta family \citep{Mothe-Diniz2005}. 
The size of the asteroid is 5.17 km in diameter \citep{Masiero2011}. 
The period of this asteroid was not known prior to this study. 
Alupka was observed during its apparition in November 2004 (Fig. \ref{fig:2508}) and yielded a rotational period of 17.70 $\pm$ 0.05 h.
The asteroid is evaluated with U = 2$\mathrm{-}$.

\bigskip
\subsubsection{(3657) Ermolova}
\begin{figure}
  \begin{center}
    \FigureFile(80mm,80mm){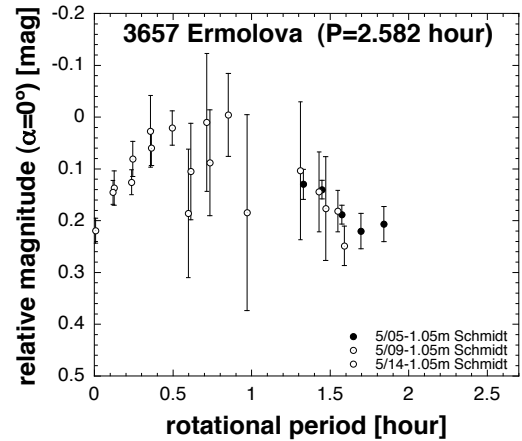}
  \end{center}
  \caption{Composite rotational lightcurve of the asteroid (3657) Ermolova.
}
\label{fig:3657}
\end{figure}

The asteroid (3657) Ermolova is a Vestoid \citep{Xu1995} and part of the Vesta family \citep{Mothe-Diniz2005}. 
Thermal observations of the asteroid have indicated it has an effective diameter of 5.79 km \citep{Usui2011}. 
Ermolova was observed during its apparition in May 2005 (Fig. \ref{fig:3657}), and a rotational period of 2.582 $\pm$ 0.004 h was determined for this asteroid. 
On the websites of Behrend and Pravec et al. (http://www.asu.cas.cz/$\sim$ppravec/neo.htm) rotational periods of 2.6066 and 2.6071 h were reported for (3657) Ermolova, respectively. 
Even taking into account uncertainties, our rotational period result is slightly different to these web-published results.
The asteroid is valued as U = 2$\mathrm{-}$.

\bigskip
\subsubsection{(3900) Knezevic}
\begin{figure}
  \begin{center}
    \FigureFile(80mm,80mm){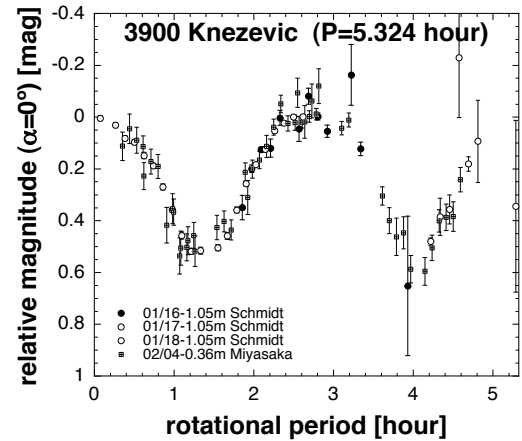}
  \end{center}
  \caption{Composite rotational lightcurve of the asteroid (3900) Knezevic.
}
\label{fig:3900}
\end{figure}

The asteroid (3900) Knezevic is a V-type asteroid \citep{Bus2002} of the Vesta family \citep{Mothe-Diniz2005}. 
The asteroid is 4.96 km in diameter \citep{Masiero2011}. 
The period of this asteroid was not known prior to our study. Knezevic was observed during its apparition in January and February 2005 (Fig. \ref{fig:3900}) and gave a rotational period of 5.324 $\pm$ 0.001 h.
The asteroid is rated as U = 3$\mathrm{-}$.

\bigskip
\subsubsection{(4005) Dyagilev}
\begin{figure}
  \begin{center}
    \FigureFile(80mm,80mm){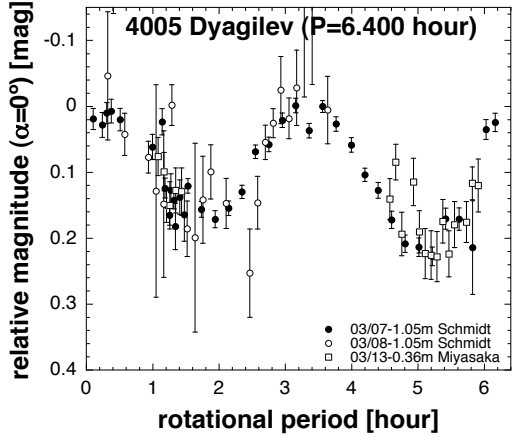}
  \end{center}
  \caption{Composite rotational lightcurve of the asteroid (4005) Dyagilev.
}
\label{fig:4005}
\end{figure}

The asteroid (4005) Dyagilev is a V-type asteroid \citep{Xu1995}, but is not a member of Vesta family \citep{Mothe-Diniz2005}. 
However, the asteroid is positioned in the inner main belt region. 
\citet{Masiero2011} estimated that the diameter of the asteroid (4005) Dyagilev is 8.36 km. 
The period of this asteroid was not known before this study. 
Dyagilev was observed during its apparition in March 2005 (Fig. \ref{fig:4005}) and yielded a rotational period of 6.400 $\pm$ 0.006 h.
The asteroid is assessed at U = 3$\mathrm{-}$.

\bigskip
\subsubsection{(4383) Suruga}
\begin{figure}
  \begin{center}
    \FigureFile(80mm,80mm){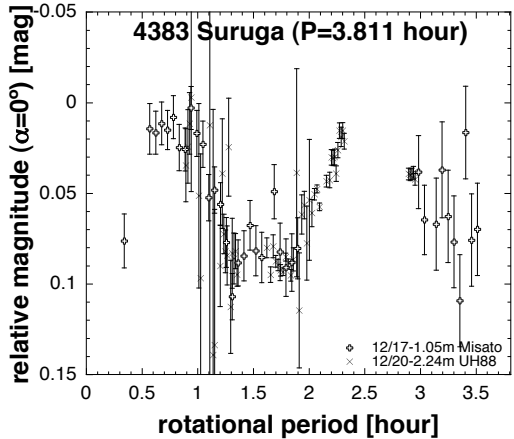}
  \end{center}
  \caption{Composite rotational lightcurve of the asteroid (4383) Suruga.
}
\label{fig:4383}
\end{figure}

The asteroid (4383) Suruga has a Vesta-like spectrum \citep{Roig2006}, but is not a member of the Vesta family \citep{Mothe-Diniz2005}, although it is positioned in the inner main belt region. 
The diameter of this asteroid has been estimated to be 7.92 km \citep{Masiero2011}. 
The period of this asteroid was not known before this study. 
Suruga was observed during its apparition in December 2004 (Fig. \ref{fig:4383}) and a rotational period of 3.811 $\pm$ 0.003 h was obtained for this asteroid.
The asteroid is estimated as U = 2$\mathrm{-}$.

\bigskip
\subsubsection{(4434) Nikulin}
\begin{figure}
  \begin{center}
    \FigureFile(80mm,80mm){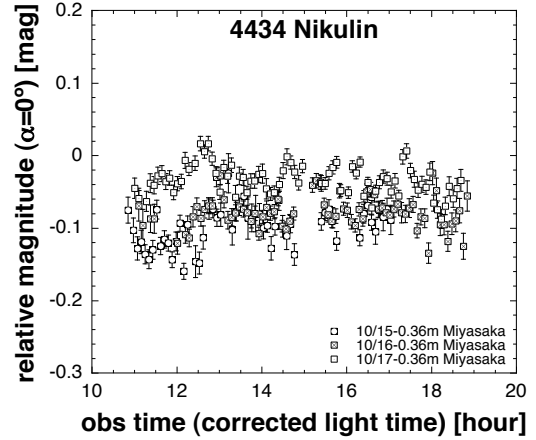}
  \end{center}
  \caption{Composite rotational lightcurve of the asteroid (4434) Nikulin.
}
\label{fig:4434}
\end{figure}

The asteroid (4434) Nikulin has a Vesta-like spectrum \citep{Xu1995}, but does not belong to the Vesta family \citep{Mothe-Diniz2005}. 
However, this asteroid is positioned in the inner main belt region. 
The asteroid is 5.11 km in diameter \citep{Masiero2011}. 
The period of (4434) Nikulin was not known before this study. Nikulin was observed during its apparition in October 2004 (Fig. \ref{fig:4434}). 
Despite three nights of observations, the rotational period of (4434) Nikulin was not able to be determined due to its small amplitude.

\bigskip
\subsubsection{(4796) Lewis}
\begin{figure}
  \begin{center}
    \FigureFile(80mm,80mm){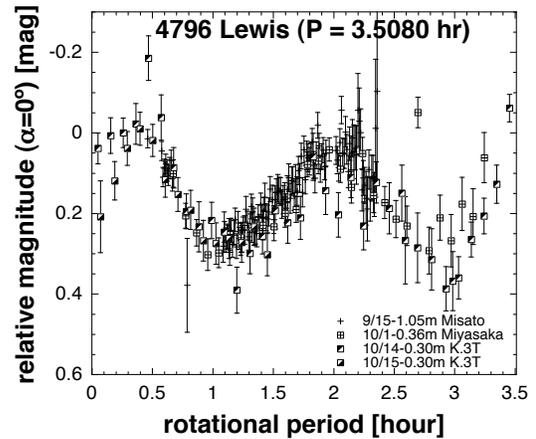}
  \end{center}
  \caption{Composite rotational lightcurve of the asteroid (4796) Lewis.
}
\label{fig:4796}
\end{figure}

The asteroid (4796) Lewis is a V-type asteroid \citep{Bus2002}, but is not a member of the Vesta family \citep{Mothe-Diniz2005}. 
However, the asteroid is positioned in the inner main belt region. 
The diameter of (4796) Lewis has been estimated to be $\sim$ 5.2 km, using the assumption that its albedo is the same as the median value of other V-type asteroids (pv = 0.343; \cite{Mainzer2012}). 
The period of this asteroid was not known before this study. Lewis was observed during its apparition in September and October 2004 (Fig. \ref{fig:4796}) and determined to have a rotational period of 3.5080 $\pm$ 0.0002 h.
The asteroid is evaluated with U = 2+.

\bigskip
\subsubsection{(6331) 1992 $\mathrm{FZ_{1}}$}
\begin{figure}
  \begin{center}
    \FigureFile(80mm,80mm){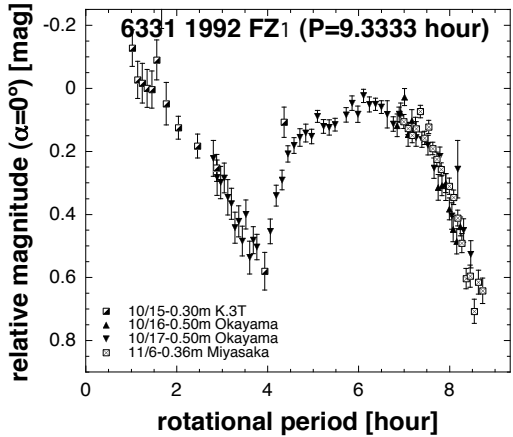}
  \end{center}
  \caption{Composite rotational lightcurve of the asteroid (6331) 1992 $\mathrm{FZ_{1}}$.
}
\label{fig:6331}
\end{figure}

The asteroid (6331) 1992 $\mathrm{FZ_{1}}$ is a Vestoid \citep{Duffard2004} that belongs to the Vesta family \citep{Mothe-Diniz2005}. 
The asteroid has an effective diameter of 5.32 km \citep{Masiero2011}. 
The period of this asteroid was not known before this study. 
1992 $\mathrm{FZ_{1}}$ was observed during its apparition in October and November 2004 (Fig. \ref{fig:6331}) and yielded a rotational period of 9.3333 $\pm$ 0.0005 h.
The asteroid is assessed at U = 2+.

\bigskip
\subsubsection{(8645) 1988 TN}
\begin{figure}
  \begin{center}
    \FigureFile(80mm,80mm){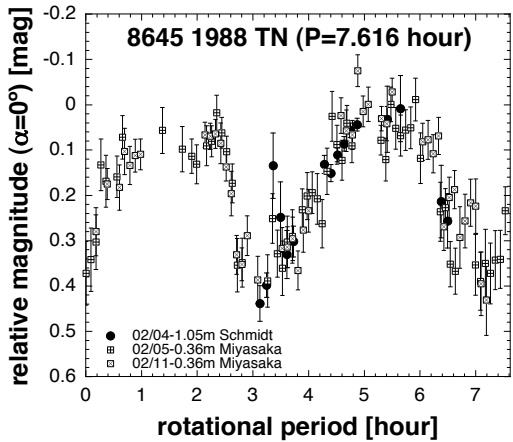}
  \end{center}
  \caption{Composite rotational lightcurve of the asteroid (8645) 1988 TN.
}
\label{fig:8645}
\end{figure}

The asteroid (8645) 1988 TN is a V-type asteroid \citep{Roig2006}, but is not a member of the Vesta family \citep{Mothe-Diniz2005}. 
However, the asteroid is positioned in the inner main belt region. 
\citet{Masiero2011} determined the diameter of this asteroid to be 5.08 km. 
The period of this asteroid was not known before this study. 
1988 TN was observed during its apparition in February 2005 (Fig. \ref{fig:8645}) and a rotational period of 7.616 $\pm$ 0.002 h determined for this asteroid.
The asteroid is rated as U = 3$\mathrm{-}$.

\bigskip
\subsubsection{(10285) Renemichelsen}
\begin{figure}
  \begin{center}
    \FigureFile(80mm,80mm){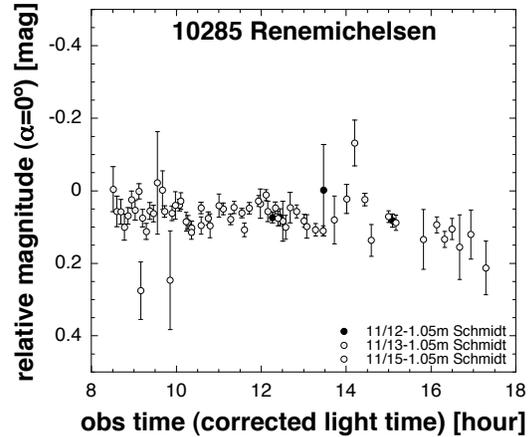}
  \end{center}
  \caption{Composite rotational lightcurve of the asteroid (10285) Renemichelsen.
}
\label{fig:10285}
\end{figure}

The asteroid (10285) Renemichelsen has a Vesta-like spectrum \citep{Duffard2004} and is a member of the Vesta family \citep{Mothe-Diniz2005}. 
Thermal observations of this asteroid have indicated that it has an effective diameter of 3.91 km \citep{Masiero2011}. 
The period of this asteroid was not known before this study. 
Renemichelsen was observed during its apparition in November 2004 (Fig. \ref{fig:10285}). 
Although (10285) Renemichelsen was observed for three nights, determination of its rotational period was not possible due to its small amplitude.

\bigskip
\subsubsection{(10320) Reiland}
The asteroid (10320) Reiland is a Vestoid \citep{Duffard2004} and a member of the Vesta family \citep{Mothe-Diniz2005}. 
The diameter of this asteroid is 2.86 km \citep{Masiero2011},  
but the period of this asteroid was not known prior to this study. 
Reiland was observed during its apparition in November 2004 (Fig. \ref{fig:8645}), and a rotational period of 5.92 $\pm$ 0.01 h was determined.
The asteroid is valued as U = 2$\mathrm{-}$.

\begin{figure}
  \begin{center}
    \FigureFile(80mm,80mm){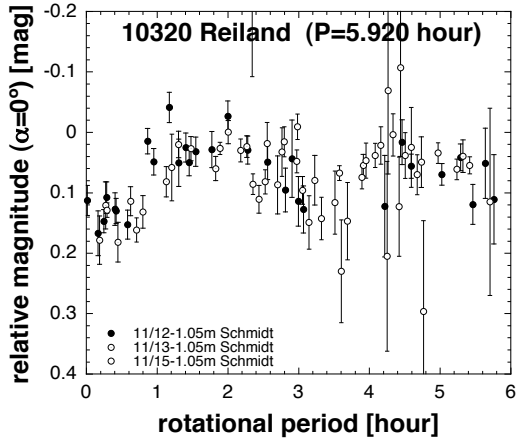}
  \end{center}
  \caption{Composite rotational lightcurve of the asteroid (10320) Reiland.
}
\label{fig:10320}
\end{figure}

\subsection{non-V-type asteroids}
Lightcurves of some non-V-type asteroids were obtained whilst making observations of V-type asteroids.

\bigskip
\subsubsection{(477) Italia}
\begin{figure}
  \begin{center}
    \FigureFile(80mm,80mm){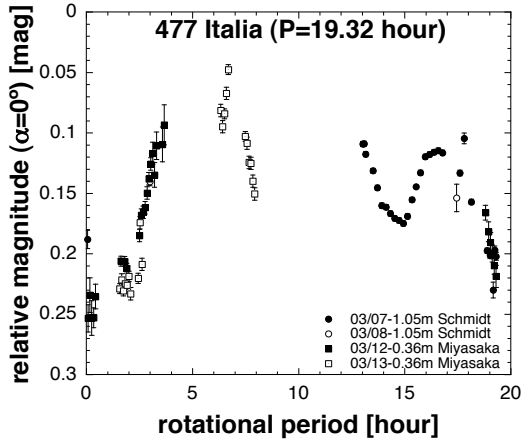}
  \end{center}
  \caption{Composite rotational lightcurve of the asteroid (477) Italia.
}
\label{fig:477}
\end{figure}

The asteroid (477) Italia is a S-type asteroid (\cite{Tholen1984}, \cite{Bus2002}) with a diameter of 25.02 km \citep{Usui2011}. 
This asteroid was observed in the same frame as (1933) Tinchen for four nights. 
The period of this asteroid was known prior to this study. 
Italia was observed during its apparition in 2004 (Fig. \ref{fig:477}) and a rotational period of 19.32 $\pm$ 0.01 h obtained. 
On the website of Behrend, the rotational period for this asteroid is reported to be 19.42 h, which is consistent with our results.
The asteroid is estimated as U = 2$\mathrm{-}$.

\bigskip
\subsubsection{(10389) Robmanning}
\begin{figure}
  \begin{center}
    \FigureFile(80mm,80mm){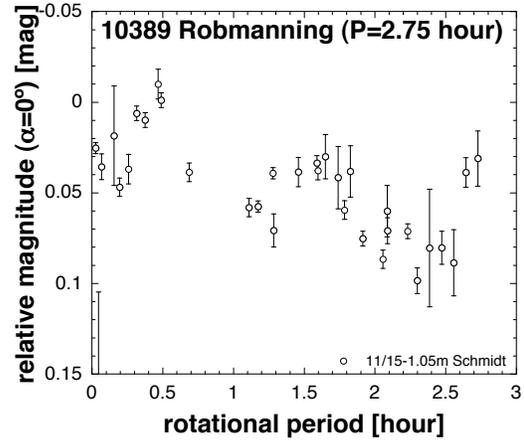}
  \end{center}
  \caption{Composite rotational lightcurve of the asteroid (10389) Robmanning.
}
\label{fig:10389}
\end{figure}

The spectral type of asteroid (10389) Robmanning is unknown. 
The diameter of the asteroid has been estimated to be 4.12 km \citep{Masiero2011}. 
This asteroid was monitored in the same frame as (11320) Reiland for one night.
The period of this asteroid was not known before this study. 
Robmanning was observed during its apparition in March 2004 (Fig. \ref{fig:10389}) and yielded a rotational period of 2.75 h.
The asteroid is evaluated with U = 2$\mathrm{-}$.

\bigskip
\subsubsection{(10443) van de Pol}
\begin{figure}
  \begin{center}
    \FigureFile(80mm,80mm){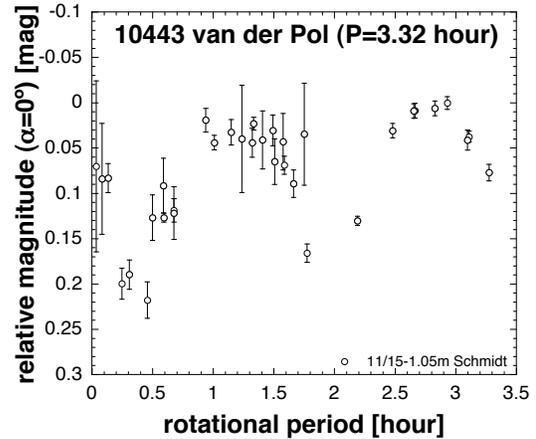}
  \end{center}
  \caption{Composite rotational lightcurve of the asteroid (10443) van de Pol.
}
\label{fig:10443}
\end{figure}

The asteroid (10443) van de Pol is an S-type asteroid \citep{Ivezic2002}. 
\citet{Masiero2011} reported that the diameter of this asteroid is 3.44 km. 
The asteroid was observed in the same frame as (11320) Reiland for one night. 
van de Pol was observed during its apparition in November 2004 (Fig. \ref{fig:10443}) and gave a rotational period of 3.32 h. 
The website of Pravec et al. reports a rotational period of 3.34501 h for this asteroid, which is consistent with our result.
The asteroid is rated as U = 2$\mathrm{-}$.

\bigskip
\subsubsection{(11321) Tosimatsumoto}
\begin{figure}
  \begin{center}
    \FigureFile(80mm,80mm){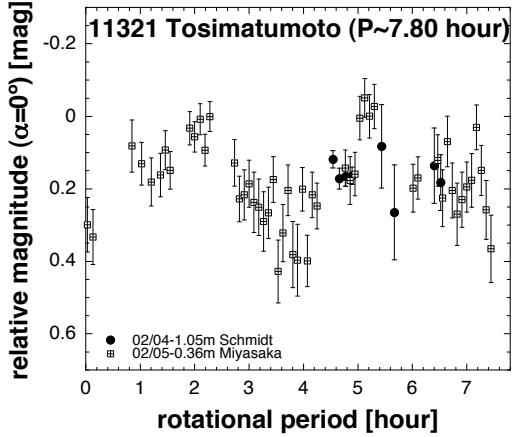}
  \end{center}
  \caption{Composite rotational lightcurve of the asteroid (11321) Tosimatsumoto.
}
\label{fig:11321}
\end{figure}

The spectrum of asteroid (11321) Tosimatsumoto classifies it as XL-type \citep{Carvano2010}, which has a diameter of 9.19 km \citep{Masiero2011}. 
This asteroid was monitored in the same frame as (8645) 1988 TN for two nights. 
The period of this asteroid was not known before this study. 
Tosimatsumoto was observed during its apparition in February 2005 (Fig. \ref{fig:11321}) and a rotational period of 7.80 $\pm$ 0.09 h was obtained.
The asteroid is assessed at U = 1+.

\bigskip
\subsubsection{(18590) 1997 $\mathrm{YO_{10}}$}
\begin{figure}
  \begin{center}
    \FigureFile(80mm,80mm){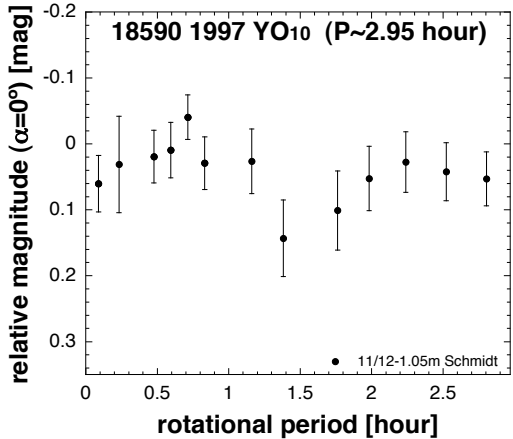}
  \end{center}
  \caption{Composite rotational lightcurve of the asteroid (18590) 1997 $\mathrm{YO_{10}}$
}
\label{fig:18590}
\end{figure}

The asteroid (18590) 1997 $\mathrm{YO_{10}}$ is an S-type asteroid (\cite{Ivezic2002}; \cite{Carvano2010}). 
Thermal observations of this asteroid have shown that it has an effective diameter of 5.02 km \citep{Masiero2011}.
The asteroid was monitored in the same frame as (2011) Veteraniya for one night. 
The period of this asteroid was not known prior to this study. 
1997 $\mathrm{YO_{10}}$ was observed during its apparition in November 2004 (Fig. \ref{fig:18590}), yielding a rotational period of $\sim$2.95 h.
The asteroid is evaluated with U = 1.

\bigskip
\subsubsection{(22034) 1999 $\mathrm{XL_{168}}$}
\begin{figure}
  \begin{center}
    \FigureFile(80mm,80mm){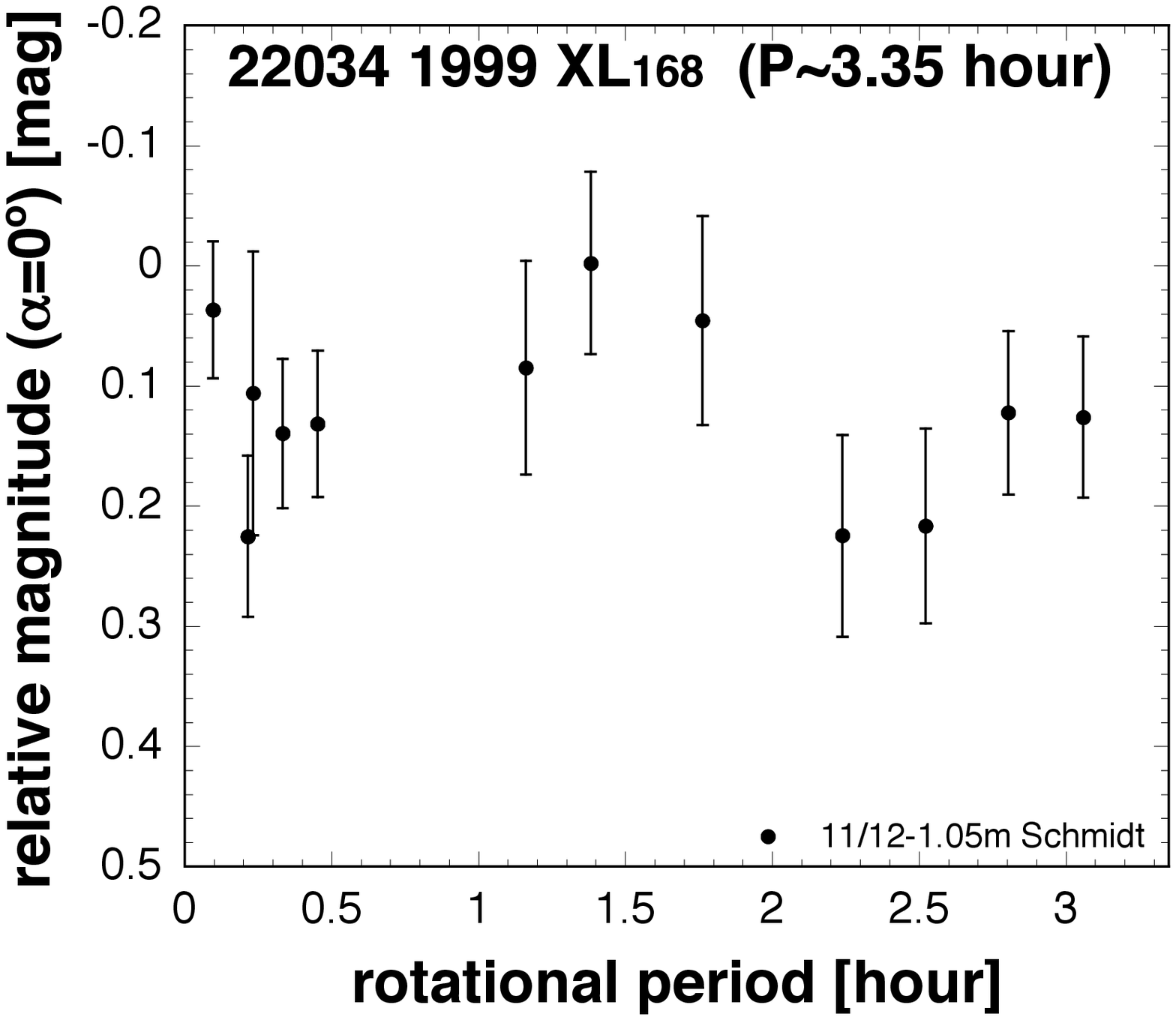}
  \end{center}
  \caption{Composite rotational lightcurve of the asteroid (22034) 1999 $\mathrm{XL_{168}}$
}
\label{fig:22034}
\end{figure}

The spectral type of asteroid (22034) 1999 $\mathrm{XL_{168}}$ is unknown. 
This asteroid has an effective diameter of 5.15 km \citep{Masiero2011}.
(22034) 1999 $\mathrm{XL_{168}}$ was monitored in the same frame as (2011) Veteraniya for one night. 
The period of this asteroid was not known prior to this study. 
1999 $\mathrm{XL_{168}}$ was observed during its apparition in November 2004 (Fig. \ref{fig:22034}), and resulted in a calculated rotational period of $\sim$3.35 h.
The asteroid is estimated as U = 1.

\bigskip
\subsubsection{(29976) 1999 $\mathrm{NE_{6}}$}
\begin{figure}
  \begin{center}
    \FigureFile(80mm,80mm){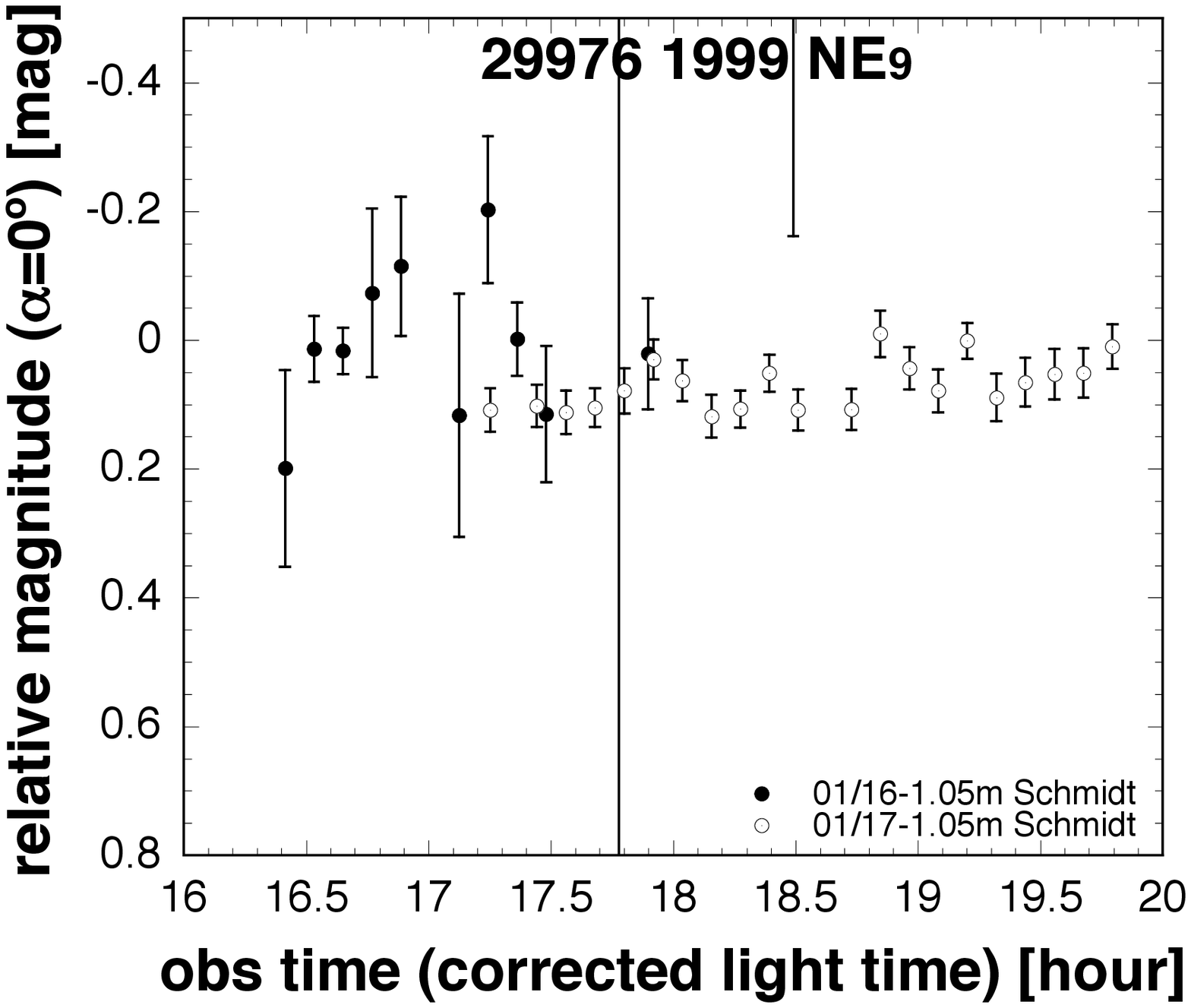}
  \end{center}
  \caption{Composite rotational lightcurve of the asteroid (22034) 1999 $\mathrm{XL_{168}}$
}
\label{fig:29976}
\end{figure}

The asteroid (29976) 1999 $\mathrm{NE_{6}}$ is a D-type asteroid \citep{Carvano2010}. 
The diameter of this asteroid has been estimated to be 9.56 km \citep{Masiero2011}. 
This asteroid was observed in the same frame as (3900) Knezevic for two nights. 
The period of this asteroid was not known before this study. 
1999 $\mathrm{NE_{6}}$ was observed during its apparition in January 2005 (Fig. \ref{fig:29976}). 
Despite two nights of observations, a solution for the rotational period of this asteroid was not possible due to its small amplitude.

\bigskip
\subsubsection{(41051) 1999 $\mathrm{VR_{10}}$}
\begin{figure}
  \begin{center}
    \FigureFile(80mm,80mm){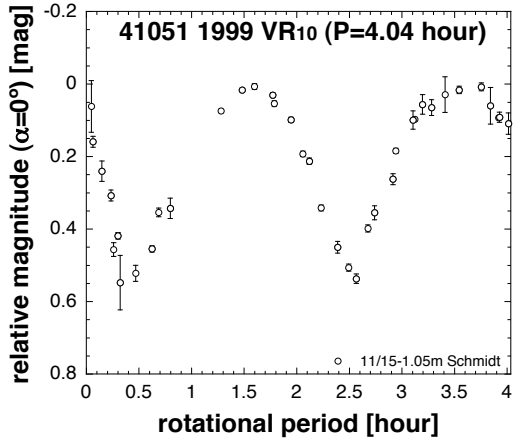}
  \end{center}
  \caption{Composite rotational lightcurve of the asteroid (41051) 1999 $\mathrm{VR_{10}}$
}
\label{fig:41051}
\end{figure}

The spectral type of asteroid (41051) $\mathrm{VR_{10}}$ is unknown. 
The diameter of the asteroid has been estimated to be 11.12 km \citep{Masiero2011}.
The asteroid was monitored in the same frame as (11320) Reiland for one night. 
The period of this asteroid was not known prior to this study. 
1999 $\mathrm{VR_{10}}$ was observed during its apparition in November 2004 (Fig. \ref{fig:41051}), yielding a rotational period of 4.04 h.
The asteroid is evaluated with U = 2$\mathrm{-}$.

\bigskip
\subsubsection{(46121) 2001 $\mathrm{FB_{36}}$}
\begin{figure}
  \begin{center}
    \FigureFile(80mm,80mm){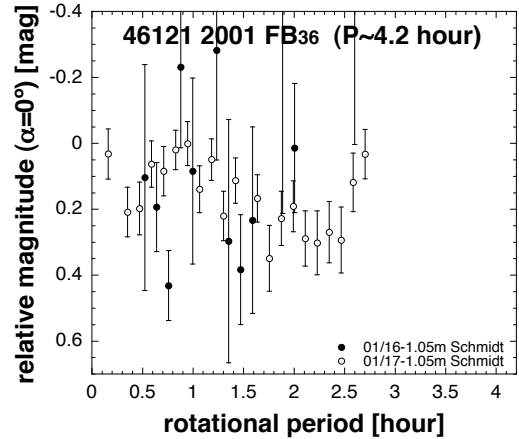}
  \end{center}
  \caption{Composite rotational lightcurve of the asteroid (46121) 2001 $\mathrm{FB_{36}}$
}
\label{fig:46121}
\end{figure}

The asteroid (46121) 2001 $\mathrm{FB_{36}}$ is a C-type asteroid (\cite{Ivezic2002}; \cite{Carvano2010}). 
\citet{Masiero2011} reported that the diameter of this asteroid is 6.26 km.
This asteroid was observed in the same frame as (3900) Knezevic for two nights. 
The period of this asteroid was not known prior to this study. 
2001 $\mathrm{FB_{36}}$ was observed during its apparition in January 2005 (Fig. \ref{fig:46121}) and a rotational period of $\sim$4.2 h was determined.
The asteroid is assessed at U = 1.

\bigskip
\subsubsection{(89481) 2001 $\mathrm{XH_{27}}$}
\begin{figure}
  \begin{center}
    \FigureFile(80mm,80mm){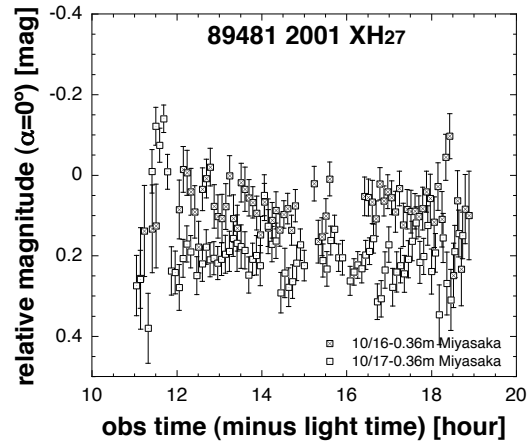}
  \end{center}
  \caption{Composite rotational lightcurve of the asteroid (89481) 2001 $\mathrm{XH_{27}}$
}
\label{fig:89481}
\end{figure}

(89481) $\mathrm{XH_{27}}$ is an X-type asteroid \citep{Carvano2010} with 
an albedo that is the same as the median value of X-type asteroids \citep{Mainzer2012}. The size of this asteroid is $\sim$ 2.8 km. 
The Jupiter Trojan asteroid was monitored in the same frame as (4434) Nikulin for two nights.
The period of this asteroid was not known before this study. 2001 $\mathrm{XH_{27}}$ was observed during its apparition in October 2005 (Fig. \ref{fig:89481}). 
Although observed for two nights, its rotational period was not able to be determined due to its small amplitude.

\section{Discussion}
We have now determined the rotational rate of 59 V-type asteroids in the inner main belt, which includes our 11 new and 48 previous results (Table \ref{tab:V-type asteroids}). 
The diameters of these asteroids range from 3 to 10 km, respectively, on the basis of an assumed albedo of the median value of V-type asteroids. 

\renewcommand{\tabcolsep}{3pt} 
\begin{longtable}{rlcccccccl}
  \caption{Rotational properties and orbital elements for V-type asteroids in the inner main belt.}\label{tab:V-type asteroids}
  \hline              
  NUM & NAME & Spin rate & Da\footnotemark[$*$] & a  & e & i & Spectrum & Vesta & Lightcurve
\\ 
       &     & [rev/day] & [km]            & [AU] &   & [$\timeform{D}$]  & refference\footnotemark[$\dagger$] & family & literature
\\ 
\endfirsthead
  \hline
  NUM & NAME & rate & Da & a [AU] & e & i [$\timeform{D}$] & spectrum & family & literature
\\
  \hline
\endhead
  \hline
\endfoot
  \hline
\multicolumn{1}{@{}l}{\rlap{\parbox[t]{1.0\textwidth}{\small
\footnotemark[$*$] References for diameter. \footnotemark[AF]\citet{Usui2011}; \footnotemark[WS]\citet{Masiero2011}; \footnotemark[W3]\citet{Masiero2012}; \footnotemark[KA]assuming albedo is the same as the median value of V-type asteroids (pv = 0.343; \cite{Mainzer2012}).
\\
\footnotemark[$\dagger$] References for spectral type. AC: \citet{Alvarez-Candal2006}; BP: Binzel, R.P. private comm. CA: \citet{Carvano2010}; D4: \citet{Duffard2004}; D6: \citet{Duffard2006}; RG: \citet{Roig2006}; SM1: \citet{Xu1995}; SM2: \citet{Bus2002}; S3: \citet{Lazzaro2004}.
}}}
\endlastfoot
  \hline
809&Lundia&1.56 &10.26\footnotemark[WS] &2.2829 &0.1925 &7.14 &S3,D4& &\citet{Kryszczynska2009} \\
854&Frostia&0.64 &9.49\footnotemark[AF] &2.3682 &0.1740 &6.09 &AC& &\citet{Behrend2004}\\
956&Elisa&6.17 &10.47\footnotemark[WS] &2.2982 &0.2038 &5.96 &S3,D4 &&Behrend, R. website\footnotemark[2]\\
1273&Helma&3.94 &10.39\footnotemark[WS] &2.3935 &0.1622 &5.42 &SM1&1&\citet{Higgins2007}\\
1906&Naef&2.18 &8.06\footnotemark[WS] &2.3732 &0.1351 &6.47 &SM1&1&\citet{Durkee2007}\\
1929&Kollaa&8.03 &7.77\footnotemark[WS] &2.3630 &0.0746 &7.78 &SM1&1&Behrend, R. website\footnotemark[2]\\
1933&Tinchen&6.54 &6.45\footnotemark[WS] &2.3527 &0.1229 &6.88 &SM1&1&This work\\
1946&Walraven&2.35 &9.21\footnotemark[WS] &2.2934 &0.2352 &8.17 &AC& &\citet{Van1933}\\
2011&Veteraniya&2.92 &5.19\footnotemark[WS] &2.3863 &0.1502 &6.20 &SM1&1&This work\\
2113&Ehrdni&1.82 &5.20\footnotemark[KA] &2.4730 &0.0970 &6.44 &SM1& &\citet{Binzel1987}\\
2442&Corbett&2.40 &8.33\footnotemark[W3] &2.3874 &0.1182 &5.09 &SM1& &Behrend, R. website\footnotemark[2]\\
2486&Metsahovi&5.39 &8.42\footnotemark[WS] &2.2687 &0.0797 &8.41 &AC& &\citet{Pikler2007}\\
2508&Alupka&1.36 &5.17\footnotemark[WS] &2.3682 &0.1276 &6.08 &SM2&1&This work\\
2511&Patterson&5.79 &7.85\footnotemark[WS] &2.2985 &0.1039 &8.05 &SM2&1&\citet{Hasegawa2008}\\
2579&Spartacus&6.60 &4.60\footnotemark[WS] &2.2104 &0.0753 &5.78 &SM2& &Warner, B.D. website\footnotemark[3]\\
2590&Mourao&1.54 &7.88\footnotemark[W4] &2.3431 &0.1173 &6.12 &SM1&1&\citet{Galad2005}\\
2640&Hallstrom&1.05 &5.70\footnotemark[KA] &2.3981 &0.0883 &6.65 &SM2,RG& &\citet{Hasegawa2008}\\
2653&Principia&4.35 &9.88\footnotemark[W3] &2.4444 &0.0800 &4.74 &SM2& &\citet{Hasegawa2008}\\
2763&Jeans&3.08 &7.52\footnotemark[WS] &2.4049 &0.2174 &3.54 &SM2,D4& &\citet{Stephens2013}\\
2768&Gorky&5.33 &10.79\footnotemark[WS] &2.2347 &0.1713 &6.28 &AC& &\citet{Pray2008}\\
2795&Lepage&0.40 &6.14\footnotemark[W3] &2.2955 &0.0283 &6.03 &SM2& &\citet{Hasegawa2008}\\
2851&Harbin&4.43 &8.84\footnotemark[W3] &2.4791 &0.1229 &8.55 &SM2,D4& &\citet{Albers2010}\\
2912&Lapalma&4.20 &6.52\footnotemark[W3] &2.2896 &0.0707 &7.28 &SM2& &\citet{Brinsfield2008}\\
3155&Lee&2.89 &6.85\footnotemark[WS]&2.3425 &0.1011 &7.20 &SM1,SM2,D4&1&\citet{Warner2003}\\
3268&De Sanctis&1.41 &6.03\footnotemark[WS] &2.3472 &0.1264 &6.35 &SM1,D4&1&\citet{Wisniewski1991}\\
3307&Athabasca&4.90 &3.63\footnotemark[W3] &2.2593 &0.0955 &6.37 &SM2&1&\citet{Hasegawa2008}\\
3376&Armandhammer&3.03 &8.17\footnotemark[WS] &2.3487 &0.0661 &6.35 &SM2&1&Pravec, P. et al. website\footnotemark[4]\\
3536&Schleicher&4.15 &3.64\footnotemark[WS] &2.3431 &0.0492 &6.56 &SM2& &\citet{Binzel1992}\\
3657&Ermolova&9.30 &5.80\footnotemark[AF] &2.3124 &0.1324 &5.79 &SM1&1&This work\\
3703&Volkonskaya&7.42 &3.73\footnotemark[WS] &2.3319 &0.1340 &6.74 &BP&1&\citet{Ryan2004a}\\
3782&Celle&6.25 &5.92\footnotemark[WS] &2.4148 &0.0943 &5.25 &SM2,RG&1&\citet{Ryan2004b}\\
3850&Peltier&9.88 &3.77\footnotemark[KA] &2.2343 &0.1621 &5.27 &SM2& &\citet{Oey2007}\\
3900&Knezevic&4.51 &4.96\footnotemark[WS] &2.3707 &0.1376 &6.73 &SM2&1&This work\\
3968&Koptelov&1.15 &5.97\footnotemark[KA] &2.3212 &0.0923 &6.21 &SM1&1&Behrend, R. website\footnotemark[2]\\
4005&Dyagilev&3.75 &8.36\footnotemark[WS] &2.4511 &0.1497 &6.84 &SM1& &This work\\
4147&Lennon&0.18 &5.17\footnotemark[WS] &2.3617 &0.0802 &5.74 &SM1&1&\citet{Hasegawa2008}\\
4215&Kamo&1.91 &6.55\footnotemark[W3] &2.4175 &0.0619 &5.75 &SM2&1&\citet{Wetterer1999}\\
4383&Suruga&6.30 &6.47\footnotemark[WS] &2.4244 &0.0630 &7.15 &RG& &This work\\
4796&Lewis&6.84 &5.20\footnotemark[KA] &2.3553 &0.1803 &2.27 &SM2,D4& &This work\\
4977&Rauthgundis&0.39 &3.86\footnotemark[WS] &2.2924 &0.1113 &5.88 &SM2& &\citet{Hasegawa2008}\\
5111&Jacliff&8.45 &6.69\footnotemark[WS] &2.3542 &0.1266 &5.81 &RG&1&Behrend, R. website\footnotemark[2]\\
5240&Kwasan&4.49 &7.27\footnotemark[WS] &2.3831 &0.1004 &5.61 &SM2&1&\citet{Ivanova2002}\\
5481&Kiuchi&6.63 &5.70\footnotemark[KA] &2.3400 &0.0625 &5.96 &S3,RG&1&\citet{Kusnirak2008}\\
5524&Lecacheux&2.85 &19.9\footnotemark[W3] &2.3662 &0.0284 &7.49 &CA& &Pravec, P. et al. website\footnotemark[4]\\
5560&Amytis&3.11 &4.70\footnotemark[WS] &2.2856 &0.1084 &5.62 &RG& &\citet{Hawkins2008}\\
5599&1991 $\mathrm{SG_{1}}$&6.63 &7.17\footnotemark[WS] &2.4195 &0.1337 &6.46 &RG& &\citet{Willis2004}\\
5754&1992 $\mathrm{FR_{2}}$&2.70 &5.70\footnotemark[WS] &2.2676 &0.1416 &5.54 &CA& &Behrend, R. website\footnotemark[2]\\
5875&Kuga&4.32 &7.47\footnotemark[WS] &2.3792 &0.0492 &6.47 &CA& &\citet{Carbo2009}\\
6159&1991 YH&2.26 &5.34\footnotemark[WS] &2.2916 &0.0618 &6.86 &D4&1&\citet{Warner2006}\\
6331&1992 $\mathrm{FZ_{1}}$&2.57 &5.32\footnotemark[WS] &2.3585 &0.1338 &7.76 &D4&1&This work\\
6406&1992 MJ&3.52 &4.06\footnotemark[W3] &2.2758 &0.1774 &8.17 &AC& &\citet{Higgins2007}\\
6877&Giada&5.75 &4.66\footnotemark[WS] &2.3231 &0.1344 &6.18 &RG&1&Behrend, R. website\footnotemark[2]\\
6976&Kanatsu&5.96 &5.50\footnotemark[WS] &2.3334 &0.1694 &8.25 &CA& &\citet{Carbo2009}\\
8645&1988 TN&3.15 &5.08\footnotemark[WS] &2.4311 &0.0608 &5.28 &RG& &This work\\
10320&Reiland&4.05 &2.86\footnotemark[WS] &2.2881 &0.1340 &6.34 &D4&1&This work\\
17035&Velichko&8.28 &4.76\footnotemark[WS] &2.4440 &0.1463 &6.24 &RG&1&Behrend, R. website\footnotemark[2]\\
19979&1989 VJ&3.17 &6.85\footnotemark[KA] &2.4588 &0.1023 &5.17 &RG&1&\citet{Han2013}\\
35062&Sakuranosyou&1.26 &3.21\footnotemark[WS] &2.3695 &0.2388 &10.51 &CA& &Pravec, P. et al. website\footnotemark[4]\\
45073&Doyanrose&7.59 &2.73\footnotemark[KA] &2.3336 &0.1479 &6.91 &CA&1&\citet{Ruthroff2011}\\
\end{longtable}
\footnotetext[2]{$\langle$http://obswww.unige.ch/$\sim$behrend/page\_cou.html$\rangle$.}
\footnotetext[3]{$\langle$http://www.minorplanetobserver.com/PDO/PDOLightcurves.htm$\rangle$.}
\footnotetext[4]{$\langle$http://www.asu.cas.cz/$\sim$ppravec/neo.htm$\rangle$.}

\begin{figure}
  \begin{center}
    \FigureFile(80mm,80mm){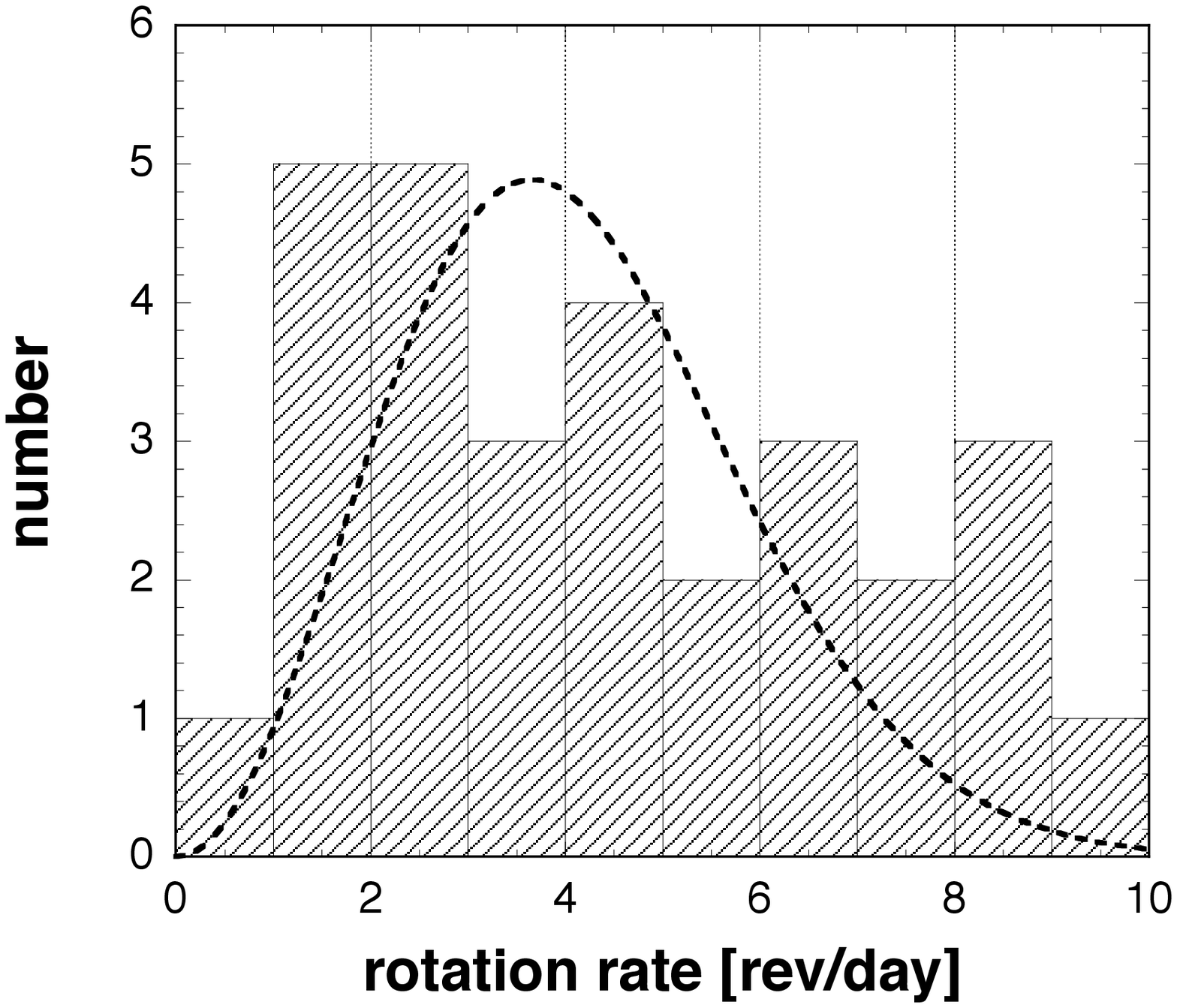}
    \FigureFile(80mm,80mm){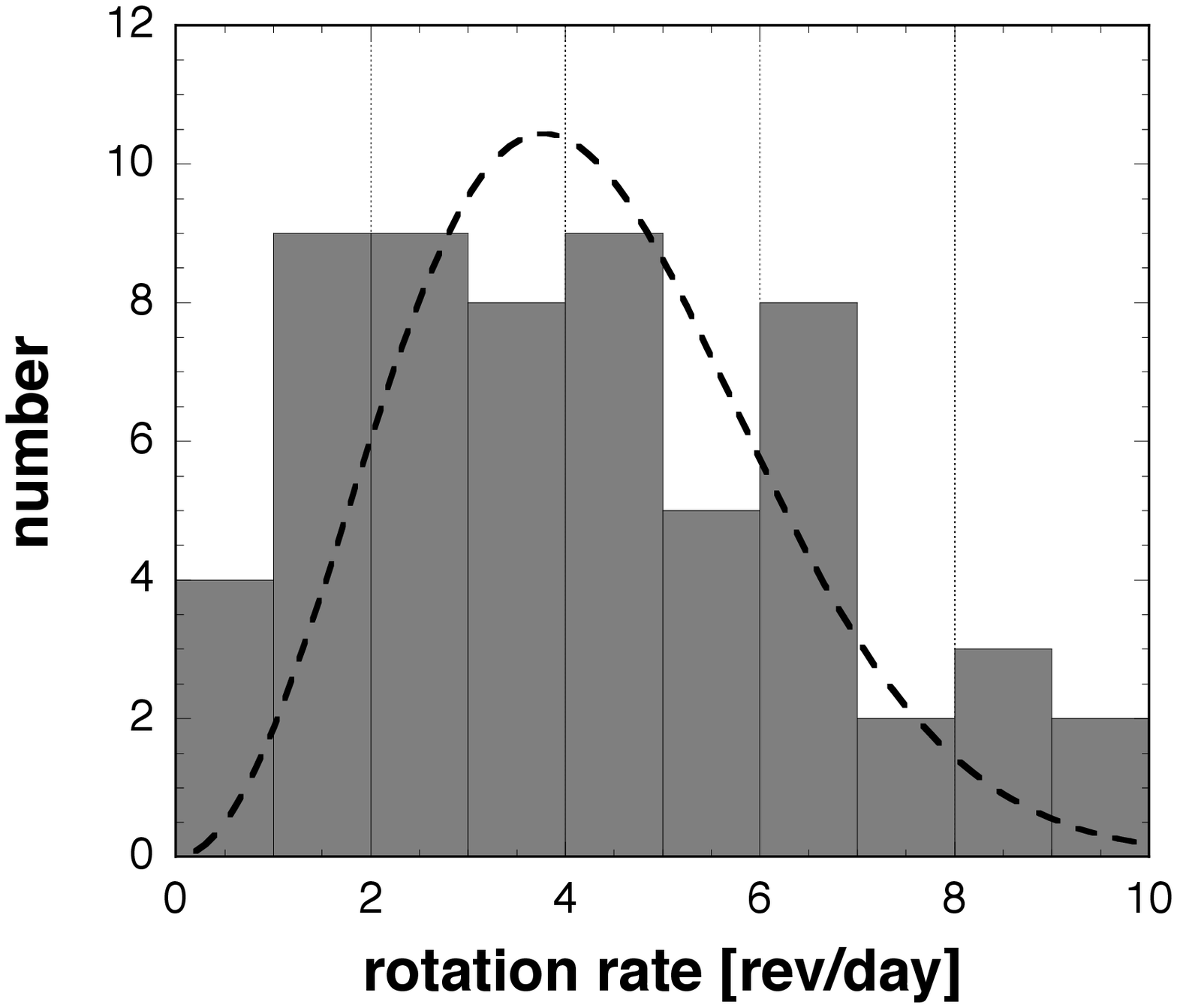}
  \end{center}
  \caption{Distribution of rotational rates for V-type asteroids in the inner main belt. (a) only V-type Vesta family asteroids. (b) all V-type asteroids. Asteroid 4 Vesta has been excluded from this figure. The dashed curve represents the best-fit Maxwellian distribution.
}
\label{fig:VD}
\end{figure}

Figure \ref{fig:VD} is a histogram of the rotational rate for the 59 V-type asteroids in the inner main belt and 29 V-type Vesta family asteroids, excluding 4 Vesta. 
Both V-type Vesta family asteroids and V-type asteroids in the inner main belt exhibit a non-Maxwellian distribution compared with the Maxwellian distribution expected for a collisionally evolved population.

There is certainly present an observational bias against slow rotators with spin rates slower than 1 rev/day and probably also against asteroids with spin rates faster than 7 rev/day. 
Measurements of rotation periods for slow rotators as well as for low amplitude ones are more difficult than for asteroids with more ordinary periods between 4 and 24 hours. 
Fast rotators with more than 7 rev/day close to the spin rate barrier that is at f about 10 rev/day tend to have spheroidal shapes and thus lower lightcurve amplitudes.  
Therefore, there become fewer data of these asteroids than real distribution.  
Taking into account observational bias, the real distributions of V-type Vesta family asteroids and V-type asteroids in the inner main belt are more different from Maxwellian distribution.

\citet{Slivan2008} showed that the distribution of rotation rates of the Koronis family is non-Maxwellian. 
\citet{Vokrouhlicky2003} suggested that the rotation rates and obliquities of the spin vector for the Koronis family have been influenced by YORP effects for several billion years. 
The timescale of YORP effects for the Koronis family is consistent with the age of the Koronis family as estimated by numerical simulations \citep{Marzari1999}, crater counting of the Koronis family asteroid 243 Ida \citep{Greenberg1996}, and age--color relationships from photometric observations \citep{Nesvorny2005}.
\citet{Kryszczynska2012} showed that the rotational rates of the Flora family also have a non-Maxwellian distribution. This result clearly demonstrates the influence of YORP effects on the Flora family. 
The age of the Flora family \citep{Nesvorny2005} is also consistent with the timescale of YORP effects.
\citet{Pravec2008} showed that the distribution of main belt and Mars-crossing asteroids with diameters of 3--15 km has a non-Maxwellian distribution due to YORP torques. 
In light of these examples of other asteroid families, it is probable that the rotational rate distribution of V-type asteroids can be explained by long-term modification by YORP torques.
The YORP effect is dependent on asteroid size, shape, spin vector, and heliocentric distance \citep{Rubincam2000}. 
As such, we speculate that the timescale for generation of V-type asteroids with 3 $\lesssim$ Diameter $\lesssim$ 10 km ranges from sub-billion to several billion years.

The results of \citet{Marzari1996} indicate that the cratering event that formed the Vesta family occurred $\sim$1 Gyr ago. 
\citet{Carruba2007} implied the Vesta family age is (1.2 $\pm$ 0.7) Gyr in age.
The Dawn spacecraft scanned the surface of Vesta and the two largest impact craters were estimated to have formed about 1 Gyr ago \citep{Marchi2012}. 
\citet{Nesvorny2008} also suggested that the Vesta family has existed for at least $\sim$1 Gyr and, if members of the Vesta family were liberated from the surface of Vesta, that the impact event occurred at 3.8 Gyr. 
Using a collisional evolution model \citet{Bottke2005}, determined with a 19\% probability that a major impact has occurred on asteroid (4) Vesta and produced the Vesta family in the last 3.5 Gyr. 
Several large impact-heating events have also been shown to have occurred on the asteroid 4 Vesta between 3.4 and 4.5 Gyr ago through Ar--Ar radiometric dating of eucrites \citep{Bogard2003}. 
In summary, the timescale for formation of the Vesta family and ejecta from Vesta, if YORP effects are invoked to explain their rotational rate distribution, is consistent with dating of HED meteorites, numerical simulations of the Vesta family, and crater counting of Vesta by the Dawn mission.

\section{Conclusions}
We have determined the rotational rate of a number of V-type asteroids in the inner main belt: (1933) Tinchen, (2011) Veteraniya, (2508) Alupka, (3657) Ermolova, (3900) Knezevic, (4005) Dyagilev, (4383) Suruga, (4434) Nikulin, (4796) Lewis, (6331) 1992 $\mathrm{FZ_{1}}$, (8645) 1998 TN, (10285) Renemichelsen, and (10320) Reiland. 
Lightcurves in the R-band of rotation periods were found for Tinchen, Veteraniya, Alupka, Ermolova, Knezevic, Dyagilev, Suruga, Lewis, 1992 $\mathrm{FZ_{1}}$, 1998 TN, and Reiland. 
A compilation of rotational rates of 59 V-type asteroids in the inner main belt shows that the rotation rate distribution is non-Maxwellian, which may be explained by the long-term effect of YORP torques.

\bigskip
We would like to express our gratitude to staff members of Kiso Observatory, Misato Observatory, and UH88 for their support during lightcurve observations. 
We are grateful to Richard Binzel and Julian Oey for sharing with us valuable information about V-type asteroids. 
This work was partially supported by the Japan Society for the Promotion of Science under a Grant-in-Aid for Scientific Research (B) 17340133.
This study was partially supported by the Space Plasma Laboratory, ISAS, and JAXA as a collaborative research program.


\end{document}